

\overfullrule=0.0pt
\input psfig.tex
\magnification=1200

\font\eightrm=cmr8
\font\tit=cmbx10 scaled\magstep 2

\font\aun=cmbx10
\font\rauth=cmr10

\font\eightmit=cmmi8
\scriptfont4=\seveni
\vsize=24.0truecm
\hsize=15.0truecm
\hoffset=0.5 truecm
\voffset=0.0 truecm
\baselineskip=16pt plus 2pt minus 1pt
\parskip=2pt minus 1pt
\parindent=1.0truecm
\tolerance=1000

\def\a {\alpha}
\def\b {\beta}
\def\m {\mu}

\def\bl {\biggl}
\def\br {\biggr}
\def\ov {\over}

\def\t {\theta}

\def\v {\varphi}
\def\l {\lambda}
\def\M {{\cal M}}
\def\ltord{\hbox{$\;\raise.4ex\hbox{$<$}\kern-.75em\lower.7ex\hbox{$\sim$}
		       \;$}}
\def\gtord{\hbox{$\;\raise.4ex\hbox{$>$}\kern-.75em\lower.7ex\hbox{$\sim$}
		       \;$}}
\def\subh{\hbox{\lower.4ex\hbox{$_h$}}}
\def\subq{\hbox{\lower.2ex\hbox{$_q$}}}
\def\subrhat{\hbox{\lower.2ex\hbox{$_{{\bf \hat {\it r}}}$}}}
\def\sub0hat{\hbox{\lower.2ex\hbox{$_{{\bf \hat {\it 0}}}$}}}

\def\references{\bigskip\medskip\goodbreak{\bf\noindent
References}\nobreak
\medskip\nobreak
\frenchspacing\pretolerance=2000\parindent=25truept}
\def\paper#1#2#3#4#5#6{\item{\hbox to 20truept{[#1]\hfill}}
{\rauth #2} {#3} {\it #4} {\bf #5}, #6\smallskip}
\def\preprint#1#2#3#4{\item{\hbox to 20truept{[#1]\hfill}}
{\rauth #2} {#3} {\it #4} \smallskip}
\def\book#1#2#3#4#5{\item{\hbox to 20truept{[#1]\hfill}} {\rauth #2}
{#3} {\it #4}, #5 \smallskip}
\def\centra#1{\vbox{\rightskip=0pt plus1fill\leftskip=0pt plus1fill #1}}
\def\title#1{\baselineskip=30truept\parindent=0pt\centra{\tit #1}
\bigskip\bigskip\bigskip\baselineskip=16pt\def\titolo{#1}}
\def\first#1#2{{\aun #1$^{#2}$}}
\def\last#1#2{ \& {\aun #1$^{#2}$}}
\def\authors#1{\bigskip\centra{#1}}
\long\def\addresses#1{\bigskip\bigskip\bigskip\bigskip\centra{#1}}
\long\def\addr#1#2{$^{#1}${#2}\par}
{\nopagenumbers
\null
\vskip 5 truecm

\title{RELATIVISTIC RADIATIVE TRANSFER FOR SPHERICAL FLOWS}
\authors{\first{Luciano Rezzolla}{1}\last{John C. Miller}{1,2,3}}
\addresses{
\addr{1}{Scuola Internazionale Superiore di Studi Avanzati, Trieste, Italy.}
\addr{2}{Department of Physics, University of Oxford, England.}
\addr{3}{Osservatorio Astronomico di Trieste, Trieste, Italy.}}
\vfill\eject
\null
\vfill\eject
}
\pageno=1
\null
\vskip 5 truecm

\centerline {{\bf Abstract}}
\bigskip
\noindent
We present a new complete set of Lagrangian relativistic hydrodynamical
equations describing the transfer of energy and momentum between a
standard fluid and a radiation fluid in a general non-stationary
spherical flow. The new set of equations has been derived for a
particular application to the study of the cosmological Quark--Hadron
transition but can also be used in other contexts.
\par

\vfill\eject

\parindent=1.0truecm

\noindent
{\bf  1. Introduction }
\bigskip

An important problem in the study of relativistic plasmas is the correct
treatment of the coupling between radiation and matter. The assumptions
of the classical theory of radiative transfer are no longer adequate
near to astrophysical compact objects or for some cosmological
situations and a relativistic theory of radiation hydrodynamics is
required for these.
\par

In this paper we derive a new system of Lagrangian relativistic
hydrodynamic equations for describing general non-stationary spherical
flows in which a transfer of energy and momentum occurs between a
standard perfect fluid and a generalized ``radiation fluid'' (i.e. a
fluid composed of any effectively zero rest-mass particles having longer
mean-free-path than those of the standard fluid). Our primary interest
is in application of these equations to cosmological phase transitions
and to spherical accretion onto black holes. This paper presents the
basic system of equations; further details of the applications will be
given in subsequent papers. \par

Consistent treatments of relativistic radiation hydrodynamics have only
been developed fairly recently and among these, particular interest has
been focussed on the PSTF tensor formalism devised by K.S. Thorne in
1981 [$1,\,2$] and the covariant flux-limited diffusion theory (FLDT)
approach proposed by A.M. Anile et al. [3]. Both methods represent
approximations to a problem which is too complicated to handle in full
detail at the present time and there has been some debate about their
respective merits. Since the PSTF method has already been used for a
number of successful calculations [$4,\,5,\,6$], we have followed this
approach here but we await with interest results from applications of
the alternative method. We will comment in subsequent sections on our
reasons for thinking that the approach which we are following is a
satisfactory one in the present context. \par

In a recent paper, M.-G. Park [7] has presented a system of relativistic
radiation hydrodynamic equations for treating a time-dependent spherical
accretion flow onto a black hole or neutron star. The approach there is
rather different from that of the present work and, in particular,
Park writes the fluid equations in an Eulerian (fixed) frame. While this
form is convenient for some purposes, the Lagrangian formulation has
great advantages for many practical applications in computations of
one-dimensional time-dependent flows (having planar, cylindrical or
spherical symmetry).

In Section 2 we briefly describe the fundamentals of the PSTF formalism
and in Sections 3 and 4 we discuss the derivation of the hydrodynamical
equations for the radiation fluid and the standard fluid. We here follow
the approach of the earlier papers [$8,\,9,\,10$] which are relevant
background for the present work. The subsequent sections introduce
further additional equations which are necessary for treating the
problem of bubble growth during the cosmological Quark--Hadron
transition. Section 5 contains a short description of the scenario for
the transition and in Section 6 we present interface junction conditions
and characteristic equations for the treatment of a discontinuity
surface between two phases of the standard fluid. Discussion of the
numerical implementation of the formalism presented here will be delayed
until a subsequent paper [11] but, as evidence that a successful
implementation can be made, some sample results are shown in
\hbox{Section 7.}
\par

In view of our application to a problem involving input from particle
physics, we use in this paper a system of units in which $c = \hbar =
k_{_B} = 1$ rather than $c = G = 1$ as is usual for calculations in
general relativity. This also has the advantage that the gravitational
source terms are clearly identified because of retention of the constant
$G$ in the equations. (Note that Thorne [$1,\,2$] adopted the slightly
different convention $ c = h = 1$). We use a space-like signature
$(-,+,+,+)$. Greek indices are taken to run from 0 to 3 and Latin
indices from 1 to 3. Covariant derivatives are denoted with a semi-colon
and partial derivatives with a comma.\par

\bigskip
\bigskip
\noindent
{\bf  2.  The PSTF Tensor Formalism }
\bigskip

The PSTF tensor formalism is a technique for solving the general
relativistic form of the {\it radiative transfer equation}, which
describes the variation of the radiation field as it propagates through
a standard fluid. The relativistic form of this is straightforward and
can be written as
$$
{d {\cal N} \over {dl}} = \Sigma  ,
\eqno (1)
$$
where ${\cal N}$ is the distribution function for the photons (a
relativistic invariant), $l$ is a non-affine parameter measuring the
proper spatial distance travelled by the photons as seen from the
standard fluid, and $\Sigma$ is a source function. Note that the total
derivative is taken not just in the space-time but rather in the
phase-space since ${\cal N} = {\cal N} (x^{\alpha}, p^{\alpha})$, where
{\bf p} is the photon four-momentum. \par

The fundamental idea of the PSTF method consists of replacing equation
(1) (which is in a concise form but embodies enormous complexity) by a
hierarchy of moment equations written in terms of Projected Symmetric
and Trace-Free (PSTF) tensors which are suitably defined at each point
in the projected tangent space to the fluid four-velocity {\bf u}. The
$k$-th moments of ${\cal N}$ and $\Sigma$ are
$$
{\cal  M }^{\ \alpha_{1} \ldots \alpha_{k}}_{\nu} =
\biggl(
\int{ {{\cal N} \, \delta (2 \pi \nu + {\bf p \cdot u)}
\over {{ {\bf (-p \cdot u)}^{k-2}}}}
p^{\alpha_{1}} \ldots p^{\alpha_{k}} d {\bf V_{p}}
}\biggr) ^{PSTF} ,
\eqno(2)
$$
$$
{\cal  S }^{\ \alpha_{1} \ldots \alpha_{k}}_{\nu} =
\biggl(
\int{ {\Sigma \, \delta (2 \pi \nu + {\bf  p \cdot u)}
\over {{ {\bf (-p \cdot u)}^{k-2}}}}
p^{\alpha_{1}} \ldots p^{\alpha_{k}} d {\bf V_{p}}
}\biggr) ^{PSTF} ,
\eqno(3)
$$
where $\nu$ is the specific frequency under consideration, $ d {\bf V_{p}} $
is the invariant \hbox{momentum-space} volume element on the light cone
and $\delta (y)$ is the Dirac delta function. In the following, we will
refer to ${\cal  M }^{\ \alpha_{1} \ldots \alpha_{k}}_{\nu}$ and
${\cal  S }^{\ \alpha_{1} \ldots \alpha_{k}}_{\nu}$ simply as the $k$-th
{\it moment} and {\it source moment} respectively.
\par

The expressions (2) and (3) can be integrated over frequency and a clear
physical interpretation can be given for the first three integrated
moments of the hierarchy: ${\cal M}$ (the zero-th moment) is the energy
density of the radiation, ${\cal M}^{\a}$ (the first moment) is the
radiative energy flux, and ${\cal M}^{\a \b}$ (the second moment) is the
shear stress tensor of the radiation fluid (each quantity being measured
in the local rest frame of the standard fluid). The stress-energy tensor for
the radiation $T_{_R}^{\a \b}$ is completely defined in terms of the first
three moments and higher order moments do not enter into this definition. The
expression for it is
$$
T^{\a \b}_{_R} = {\cal M} u^{\a} u^{\b} + 2 {\cal M}^{(\a}u^{\b)}
 + {\cal M}^{\a \b}  + {1\over 3} {\cal M} P^{\a \b} ,
\eqno(4)
$$
where $P^{\a \b} (\equiv g^{\alpha\beta} + u^{\alpha}u^{\beta})$ is the
projection operator orthogonal to {\bf u}. A consequence of this is
that, if the hierarchy is truncated at the second order, it is possible
to derive the equations governing the hydrodynamics of the radiation
fluid in a particularly simple way by starting from the conservation
laws of energy and momentum. If, on the other hand, orders higher than
the second are retained, it is necessary to make direct use of the
appropriate hierarchy of equations derived from (1).
\par

In the case of planar or spherical symmetry, the $ 2k+1 $ independent
components of each $ k $ rank tensor depend on a single scalar variable
so that the tensor formalism reduces to a purely scalar one. This
simplification has made it possible for the method to be used for a
number of astrophysical applications. The hierarchy of integrated scalar
moment equations, into which equation (1) is recast, has the property
that, for any $k$, the first $k$ equations involve the first $k + 1$
moments. In order to use this scheme for making calculations, it is
necessary to truncate the moment hierarchy at some finite order by
introducing a closure relation which specifies the value of the highest
moment used in terms of lower ones and which is derived on the basis of
physical considerations.
\par

In the next section we use the PSTF approach to derive a new set of
hydrodynamical equations describing the coupling between radiation
and matter in the case of a non-stationary spherically symmetric flow,
with the moment equation hierarchy being truncated at the second
order.\par

\bigskip
\bigskip
\noindent
{\bf  3. Hydrodynamics of the Radiation Fluid }
\bigskip

We consider a spherically symmetric flow in which a transfer of energy
and momentum occurs between a generalized ``radiation fluid'' and a
standard fluid. It is convenient to use a Lagrangian frame comoving with
the standard fluid and the spherically symmetric line element
$$
ds^2 = -a^2 dt^2 + b^2 d\mu^2 + R^2 ( d\theta^2 +
sin^2 \theta \ d\varphi^2) ,
\eqno(5)
$$
where $\mu$ is a comoving radial coordinate and $R$ is an associated
Eulerean coordinate (the Schwarzschild circumference coordinate). \par

We here describe the radiation hydrodynamics using the first two moment
equations together with a closure relation and the calculations then
involve the first three moments, together with the first two
source moments. The maximum errors in the calculated values of
the radiation variables which arise when truncating at the second order,
are typically $\sim 15\%$ [8]. In view of the fact that other
uncertainties in the specification of the problem are of a comparable
order, we regard this level of approximation as acceptable for the
present purposes. Similar degrees of accuracy are reported for
calculations using FLDT schemes [12].
\par

For spherical symmetry the first three moments can be written
as
$$
{\cal M} = w_0 ,
\eqno(6)
$$
$$
{\cal M}^{\a}=w_1 e^{\a}_{{\bf {\hat {\it r}}}} ,
\eqno (7)
$$
and
$$
{\cal M}^{\a\b}= w_2 \biggl({e^{\a}_{{\bf \hat {\it r}}}
e^{\b}_{{\bf \hat {\it r}}}-
{1\over 2} e^{\a}_{{\bf \hat \theta}} e^{\b}_{{\bf \hat \theta}} -
{1\over 2} e^{\a}_{{\bf \hat \varphi}}
e^{\b}_{{\bf \hat \varphi}}}\biggr) ,
\eqno (8)
$$
where $w_0$, $w_1$ and $w_2$ are the {\it scalar moments} and $({\bf
e}\sub0hat ,\; {\bf e}\subrhat,\; {\bf e}_{{\bf \hat \theta}},\; {\bf
e}_{{\bf \hat \varphi}})$  is the orthonormal tetrad carried by an
observer comoving with the standard fluid. \par

The quantities $w_0$, $w_1$ and $w_2$ all have direct physical
interpretations corresponding to those of the related tensor moments:
$w_0$ and $w_1$ are the energy density and flux of the radiation in the
fluid rest frame, while $w_2$ is the shear stress scalar of the
radiation. The {\it scalar source moments} $s_0$ and $s_1$,
defined in a similar way to $w_0$ and $w_1$, also have direct physical
interpretations, representing the transfer of energy and momentum
between the two fluids. We use the following expressions for these
scalar source moments
$$
s_0 = {1 \over {\lambda}} (\epsilon - w_0) + \left( s_0 \right)_{_{SC}},
\eqno(9)
$$
$$
s_1 = - {w_1 \over {\lambda}} ,
\eqno(10)
$$
where $\lambda$ is the effective mean-free-path of the radiation
particles as they move through the standard fluid, $(s_0)_{_{SC}}$ is a term
expressing the contribution due to scatterings, whose form depends on the
specific problem, and $\epsilon$ is the energy density for radiation in
thermal equilibrium with the standard fluid. Assuming that it follows a
\hbox{black-body} law, $\epsilon$ can be written as
$$
\epsilon = g_{_R} \left({\pi^2 \over {30}}\right) T_{_F}^4 ,
\eqno(11)
$$
with $g_{_R}$ being the number of degrees of freedom of the radiation fluid and
$T_{_F}$ the temperature of the standard fluid. \par

Having restricted ourselves to the use of the first three scalar moments
and the first two scalar source moments, it is convenient to derive the
radiation hydrodynamical equations by means of the standard conservation
laws for the energy and momentum of the radiation fluid applied to the
stress-energy tensor (4). Following this procedure, which is equivalent
to using the scalar moment equations, we then write the following three
radiation hydrodynamical equations

$$
-u_{\a} T^{\a \b}_{{_R}\;\; ;\b} = s_0 ,
\eqno(12)
$$
$$
n_j P_{\ \a}^j T^{\a \b}_{{_R}\;\; ;\b} = {s_1 \over {b}} ,
\eqno(13)
$$
$$
w_2 = f_{\!_E} w_0 ,
\eqno(14)
$$
\medskip\noindent
where {\bf n} is a radial spacelike unit vector normal to {\bf u}.
\par

The term $f_{\!_E}$ in the closure relation (14) is a {\it variable Eddington
factor} and indicates the degree of anisotropy of the radiation. It can
take values ranging from $0$ for complete isotropy (which could, for
example, be caused by the medium being extremely optically thick) to $2/3$
for complete anisotropy. A key point in the present technique is that an
expression for $f_{\!_E}$ has to be supplied, constructed on the basis of
physical considerations and how this is done is, to some extent, {\it ad
hoc}. However, experience has shown that as long as the expression has
the correct asymptotic behaviour in any relevant limits, results do not
normally depend sensitively on the precise form used as long as it gives
a suitably smooth join between the limits [4]. This is something which needs
to be checked in any particular application but provided that the
outcome of such a check is satisfactory, it is reasonable to proceed
with confidence. \par

Making use of the stress-energy tensor (4) and of the line element (5),
equations (12) and (13) can be written explicitly as
$$
(w_0)_{,t}+{a\over b}(w_1)_{,\mu}
+{4\over 3}\biggl({{b_{,t}\over b}+{2R_{,t}\over R}}\biggr)w_0
+{2a\over b}\biggl({{a_{,\mu}\over a}+{R_{,\mu}\over R}}\biggr)w_1
+\biggl({{b_{,t}\over b}-{R_{,t}\over R}}\biggr)w_2=as_0 ,
\eqno(15)
$$
$$
(w_1)_{,t}+{a\over b}\biggl({{1\over 3}w_0+w_2}\biggr)_{\!\! ,\mu} +
{4a_{,\mu}\over {3b}}w_0
+2\biggl({{b_{,t}\over b}+{R_{,t}\over R}}\biggr)w_1
+{a\over b}\biggl({ {a_{,\mu}\over a}+{3R_{,\mu}\over R}}\biggr)w_2
=as_1 ,
\eqno(16)
$$
(see Appendix A for details). Equations (14) -- (16) are our final form
of the hydrodynamical equations for the radiation fluid and they need to
be solved together with the corresponding hydrodynamical equations for
the combined fluids which will be discussed in the next section.

\bigskip
\bigskip
\noindent
{\bf  4. Hydrodynamics of the Combined Fluids }
\bigskip

The derivation of the hydrodynamical equations for the combined fluids
(i.e. standard fluid and radiation fluid) is, in principle, more
straightforward. We follow the conventions and notation of the
earlier papers [$9,\,10$] (except that here we always write partial
derivatives using a comma) and start from the Einstein field equations
$$
R_{\a \b} - {1 \over 2} g_{\a \b} R = 8 \pi G T_{\a \b} ,
\eqno(17)
$$
with $T^{\a \b} = T^{\a \b}_{_R} + T^{\a \b}_{_F} $ being the {\it
total} stress-energy tensor, and treat the standard fluid as perfect so that
\hbox{$T^{\a \b}_{_F} = (e+p) u^{\a}u^{\b} + p\,g^{\a\b}$}. The
four independent equations which follow from (17) can be used to write
out explicitly the conservation equations for energy and momentum of the
combined fluids and the continuity equation for the standard fluid
$$
-u_{\a} T^{\a \b}_{\ \ \ ;\b} = 0 ,
\eqno(18)
$$
$$
n_jP^j_{\ \a} T^{\a \b}_{\ \ \ ;\b} = 0 ,
\eqno(19)
$$
$$
(\rho u^{\a})_{;\a} = 0 ,
\eqno(20)
$$
where $\rho$ is the {\it relative compression factor} expressing the
variation in the proper volume of co-moving elements of the standard
fluid. (For a classical standard fluid composed of non-relativistic
particles, $\rho$ can represent the rest-mass density.) \par

After some algebra, whose relevant steps are shown in Appendix A, it is
possible to write the following set of hydrodynamical equations
$$
u_{,t}=-a\biggl[{{\Gamma\over b}\biggl({p_{,\mu}+bs_1\over {e+p}}\biggr)
+ 4\pi G R \biggl({p+{1\over 3}w_0 + w_2}\biggr)
+ {G M\over {R^2}}}\biggr] ,
\eqno(21)
$$
$$
R_{,t}=au ,
\eqno(22)
$$
$$
{(\rho R^2)_{,t}\over {\rho R^2}}=
-a\biggl({{u_{,\mu}-4\pi G b R w_1\over {R_{,\mu}}}}\biggr) ,
\eqno(23)
$$
$$
e_{,t}=w\rho_{,t}-as_0 ,
\eqno(24)
$$
$$
{(aw)_{,\mu}\over {aw}}=-{w\rho_{,\mu}-e_{,\mu}+bs_1\over {\rho w}} ,
\eqno(25)
$$
$$
M_{,\mu}=4\pi R^2 R_{,\mu}\biggl({e + w_0 +{u\over {\Gamma}}w_1}\biggr) ,
\eqno(26)
$$
$$
\Gamma=\biggl({1+u^2-{2GM\over R}}\biggr)^{\!1/2}={1\over b}R_{,\mu} ,
\eqno(27)
$$
$$
b = {1 \over { 4 \pi R^2 \rho}} .
\eqno(28)
$$
Here $u$ is the radial component of fluid four velocity in the associated
Schwarzschild (Eulerian) frame, $\Gamma$ is the general relativistic
analogue of the Lorentz factor, and $w$ is the specific enthalpy ($w =
(e+p)/ \rho$). The generalized mass function $M$ can also be
calculated using the alternative equation
$$
M_{,t}=-4\pi R^2 R_{,t}
\biggl({p + {1\over 3}w_0 +{\Gamma\over u}w_1+w_2}\biggr) .
\eqno(29)
$$

The set of hydrodynamical equations (14) -- (16) and (21) -- (28) needs
to be supplemented by an equation of state, relating the energy density
$e$, the pressure $p$ and temperature $T$ of the standard fluid. The
basic set of equations is then completed and can be used for describing
the transfer of energy and momentum between a standard fluid and a
radiation fluid in a general non-stationary relativistic flow.
\par

It is worthy of note that the gravitational source terms appearing in
the combined set of equations introduce only a very minor additional
complication; for spherical symmetry, a {\it general} relativistic
calculation is only marginally more complicated than a {\it special}
relativistic one. \par

We next turn to the introduction of the further equations necessary for
treating the problem of bubble growth during the cosmological
Quark--Hadron transition. Here, there are two different phases of the
standard fluid separated by a phase interface (the bubble surface) and
so the above system of equations (which continue to hold for the bulk of
each phase) needs to be supplemented by a treatment of the interface.
Before discussing how this is done (in Section 6), Section 5 gives a
brief introduction to the scenario within which we are working. \par

\bigskip
\bigskip
\noindent
{\bf  5.  The Cosmological Quark--Hadron Transition }
\bigskip

According to the standard Hot Big-Bang model, the early universe
experienced a succession of phase transitions and breakings of symmetry
which would have influenced the subsequent evolution. The last
transition in this succession (which includes Inflation and the Electro-Weak
transition) is usually thought to have been the Quark--Hadron transition
at which strongly interacting matter passed from being a plasma of
unconfined quarks and gluons to a plasma in which the quarks and
gluons were confined within hadrons. Since it is thought to have been the
last of the cosmological transitions, any remnants which it left behind
could have had particular significance and so it is of considerable
interest to carry out a detailed hydrodynamical study in order to
see how it would have proceeded. \par

Chronologically, the transition is estimated to have taken place about
$10 \mu$s after the Big-Bang, when the temperature of the universe was
$\approx 200$ MeV, the mean density was $\approx 10^{15}$ g cm$^{-3}$
and the horizon scale was $\approx 10$ km. At present, QCD lattice
calculations carried out to investigate the nature of the transition
seem to indicate that it may well be a continuous one [13]. However,
these calculations are extremely complicated and the results are not yet
definitive. In view of the very interesting consequences which could
arise if it turns out to be a first order ({\it discontinuous}) transition
[14], extensive investigations have been carried out into this latter
picture. The work reported here is within this scenario in the case
where the transition starts with the nucleation of hadronic bubbles
within a slightly supercooled quark-gluon plasma. The bubbles of the new
phase then grow (the hadron phase is thermodynamically favoured) and
subsequently coalesce until eventually the universe is filled with the
new phase except for remaining disconnected droplets of the quark-gluon
plasma. These then shrink and can either vanish completely, possibly
leaving a significant inhomogeneity in the baryon number density [15],
or perhaps attain a stable configuration in a new hypothetical ground
state for the strongly interacting matter ({\it strange quark matter})
[16]. \par

The present paper is associated with an ongoing research programme
[$9,\,10,\,11,\,17$] which has so far focussed on studying the hydrodynamics
of the growth of single spherical hadronic bubbles within a surrounding
quark-gluon plasma. While the phase transition directly involves only
the strongly interacting particles, a crucial role is also played by
electromagnetically interacting particles (photons, electrons, muons and
their antiparticles) and particles which interact only weakly (neutrinos
and antineutrinos). These can couple to the quark and hadronic plasmas and
provide a mechanism for long-range transport of energy and momentum on
account of having longer mean-free-paths than the strongly interacting
particles. The effect of this starts to be significant when the radius
of the growing bubble becomes comparable with the relevant
mean-free-path ($\sim 10^{3}$ fermi for the electromagnetic interaction
and $\sim 1$ cm for the weak interaction) and eventually produces an
effective total coupling between the respective components when the
radius of the bubble is much greater than the relevant mean-free-path.
\par

Because of the large difference between these mean-free-paths,
long-range transport can become important at two different stages during
the growth of a bubble. However, the effects on the hydrodynamics will
be exactly the same in each case and so we will omit any distinction
between them here. The term ``standard fluid'' will always be taken to
refer to all of those particles having mean-free-paths small compared
with the current relevant scale-length for changes in the flow of the
strongly-interacting matter. \par

As mentioned in Section 4, the hydrodynamical equations need to be
supplemented by equations of state for each of the two phases. For
simplicity, we have taken the equation of state for the hadronic matter
to be that for an ideal gas of massless point-like pions
$$
e\subh = g\subh \left({\pi^2 \over {30}}\right) T^4_h \hskip 4.0truecm
p\subh = {1 \over 3} e\subh ,
\eqno(30)
$$
where $T$ is the temperature and $g$ is the degeneracy factor (the
subscripts $h$ and $q$ indicate quantities in the hadron and quark
phases respectively). The deconfined quarks and gluons cannot be
considered as entirely free and for these we have used the
phenomenological expression given by the M.I.T. {\it Bag Model} [18]
$$
e\subq = g\subq \left({\pi^2 \over {30}}\right) T^4_q + B \hskip 4.0truecm
p\subq = g\subq \left({\pi^2 \over {90}}\right) T^4_q - B ,
\eqno(31)
$$
where a positive constant (the ``Bag'' constant $B$) is added to the
energy density and subtracted from the pressure so as to take into
account the complex effects of confinement. When the photons and
relativistic leptons are completely coupled to the strongly interacting
matter, their contribution can be included by incrementing $g\subh$ and
$g\subq$ by the relevant number of additional degrees of freedom.
\par

A final comment should be made about the expression adopted in our
calculations for the term $(s_0)_{_{SC}}$, which, as mentioned in
Section 3, represents the transfer of energy by means of scatterings.
While complete expressions for this are available for simpler
applications [4], the lack of detailed knowledge of the interaction
processes in the present context has led us to use a phenomenological
approach, expressing $(s_0)_{_{SC}}$ as equal to the absorption and
emission term, (i.e. $(\epsilon - w_0)/\lambda$), multiplied by an
adjustable coefficient ranging between zero and one. Fortunately, the
results of our calculations (which will be presented in paper [11]) turn
out not to depend sensitively on the value chosen for this coefficient.

\bigskip
\bigskip
\noindent
{\bf  6.  Treatment of the Phase Interface }
\bigskip

As long as the radius of the bubble is large compared with the
strong-interaction length scale, it is reasonable to treat the interface
between the hadron and quark phases as an exact discontinuity surface
with the variables on either side of it being linked by junction
conditions. The phase interface could, in principle, move either
supersonically (as a detonation front) or subsonically (as a
deflagration front) with respect to the medium ahead. In practice,
however, the bubbles are almost certain to expand subsonically in the
present situation (see [$9,\,19$]) and we restrict our attention to this
case here.
\par

In contrast with the situation for calculations of detonation fronts,
for a deflagration it is necessary to supplement the hydrodynamical
equations with an independent expression for the rate at which quark
matter is converted into hadronic matter, derived from considerations of
the physical processes occurring at the interface. A simple expression
for this rate was presented in [9] and setting this equal to the
hydrodynamical flux across the interface, gives the additional equation
$$
{{aw\dot \mu_{_S}}\over{4\pi R^2_{_S}(a^2-b^2\dot \mu_{_S}^2)}}
	= \left({\a \over 4}\right) g\subh
	   \left({\pi^2 \over {30}}\right) (T_q^4-T_h^4)
\eqno(32)
$$
where $\mu_{_S}$ is the interface location, $\dot \mu_{_S} =
d\mu_{_S}/dt $ and $\alpha$ is an accommodation coefficient ($0\le \a
\le 1$).\par

Also, it is very important to pay attention to a correct calculation of
the causal structure in the vicinity of the interface and it seems that
the only satisfactory way of accomplishing this, when the interface is
treated as a discontinuity, is by making use of a characteristic method
[20]. This involves rewriting the system of partial differential
equations as a system of ordinary differential equations along {\it
characteristic curves} in the space-time (which can be physically
interpreted as the worldlines of sonic perturbations in the standard
fluid and radiation fluid and the flowlines of the standard fluid).
The computational technique employed is then
to use the continuum equations of the previous sections for the bulk of
each phase and to track the interface continuously through the finite
difference grid using a characteristic method together with junction
conditions (see [$10,\,11$]). \par

For each two moment equations which are retained from the infinite
hierarchy, there are two families of corresponding characteristic curves
with associated characteristic speeds. While including a larger number
of moments in general increases accuracy, the role and relevance of the
speeds associated with moments beyond the first two is controversial. We
note that since we are using only the first two moment equations in the
present work, these difficulties do not arise here. \par

In order to write the equations in a characteristic form, it is
convenient to make use of the following equalities coming from the
constraint equations obtained from (17) and from the conservation of
momentum
$$
{b_{,t}\over b}=a{u_{,\mu}\over R_{,\mu}}
-{4\pi G ab R\over {R_{,\mu}}}w_1 ,
\eqno(33)
$$
$$
{a_{,\mu}\over a}=-{p_{,\mu}+bs_1\over {\rho w}} ,
\eqno(34)
$$
and then to rewrite equations (15), (16), (21), (23), (24) as
$$
u_{,t}+{a\Gamma\over {b\rho w}}p_{,\mu} + B_0=0 ,
\eqno(35)
$$
$$
p_{,t}+{c_s^2a\rho w\over {b\Gamma}}u_{,\mu}+B_1=0 ,
\eqno(36)
$$
$$
\rho_{,t}+{a\rho\over {b\Gamma}}u_{,\mu}+B_2=0 ,
\eqno(37)
$$
$$
(w_0)_{,t}+{a\over b}(w_1)_{,\mu}
+{a\over {b\Gamma}}\biggl({{4\over 3}+f_{\!_E}}\biggr) w_0u_{,\mu}
-{2a\over {b\rho w}}w_1p_{,\mu}+B_3=0 ,
\eqno(38)
$$
$$
(w_1)_{,t}+{a\over b}\biggl({{1\over 3}+f_{\!_E}}\biggr)(w_0)_{,\mu}
-{a\over {b\rho w}}\biggl({{4\over 3}+f_{\!_E}}\biggr)w_0p_{,\mu}
+ {2a\over {b\Gamma}}w_1u_{,\mu}+B_4=0 ,
\eqno(39)
$$
where
$$
B_0=a\biggl\{{4\pi GR\biggl[{p+\biggl({{1\over 3}+f_{\!_E}}\biggr)w_0}\biggr]+
{GM\over {R^2}}+{\Gamma\over {\rho w}}s_1}\biggr\} ,
\eqno(40)
$$
$$
B_1=ac_s^2\biggl[{s_0+\rho w\biggl({{2u\over R}-{4\pi GR\over {\Gamma}}w_1}
\biggr)}\biggr] ,
\eqno(41)
$$
$$
B_2=a\rho\biggl({{2u\over R}-{4\pi GR\over {\Gamma}}w_1}\biggr) ,
\eqno(42)
$$
$$
B_3=2a\biggl({{\Gamma\over R}-{1\over {\rho w}}s_1}\biggr)w_1+{au\over R}
\biggl({{8\over 3}-f_{\!_E}}\biggr)w_0-{4\pi GaR\over {\Gamma}}
\biggl({{4\over 3}+f_{\!_E}}\biggr)w_0w_1-as_0 ,
\eqno(43)
$$
$$
B_4=-{a\over {\rho w}}\biggl({{4\over 3}+f_{\!_E}}\biggr)w_0s_1+2a
\biggl({{u\over R}-{4\pi GR\over {\Gamma}}w_1}\biggr)w_1 +
{3a\Gamma\over R}f_{\!_E}w_0-a s_1+{a\over b}w_0(f_{\!_E})_{,\mu} .
\eqno(44)
$$
In deriving equations (35) -- (39), the relations $R_{,t} = au$ and
$R_{,\mu} = b \Gamma $ have \line{been used and we have introduced the local
sound speed in the standard fluid}
$c_s = (\partial p / \partial e)^{1/2}$. If we now define the {\it state
vector}
$$
{\bf U}=\left(\matrix{u\cr\cr
		      p\cr\cr
		   \rho\cr\cr
		    w_0\cr\cr
		    w_1\cr}\right) ,
\eqno(45)
$$
equations (35) -- (39) can be written in the symbolic form
$$
{\partial {\bf U}\over {\partial t}}+
{\cal A}{\partial {\bf U}\over {\partial \mu}}+
{\bf B}=0 ,
\eqno(46)
$$
where ${\bf B}$ is the vector whose components are given by (40) -- (44).
\par
\eject
If the expression chosen for the Eddington factor $f_{\!_E}$ is
dependent on components of the state vector {\bf U}, it is necessary to
rewrite the partial derivative of $f_{\!_E}$ in (44) in terms of the
derivatives of the component variables. In doing this the elements of
the matrix ${\cal A}$ are obviously modified. For our specific
application, the expression which we have chosen for the Eddington
factor is the following (see Appendix B for a discussion of this)
$$
f_{\!_E} = {8 u^2 / 9 \over {(1 + 4 u^2/3)}}\left({\lambda_{h,q} \over
{\lambda_{h,q} + R}}\right) .
\eqno(47)
$$
Equation (39) then becomes
$$
(w_1)_{,t}+{a\over b}\biggl({{1\over 3}+f_{\!_E}}\biggr) (w_0)_{,\mu}
-{a\over {b\rho w}}\biggl({{4\over 3}+f_{\!_E}}\biggr)w_0p_{,\mu}
+ {2a\over {b\Gamma}}(1+K)w_1u_{,\mu}+B_4=0 ,
\eqno(48)
$$
with
$$
\displaylines{
B_4=-{a\over {\rho w}}\biggl({{4\over 3}+f_{\!_E}}\biggr)w_0s_1+2a
\biggl({{u\over R}
-{4\pi GR\over {\Gamma}}w_1}\biggr)w_1 + \hfill{} \cr
\hfill{}
+ {3a\Gamma\over R}f_{\!_E}w_0-a s_1 -
{a \Gamma f_{\!_E}w_0 \over{ \l_{h,q} + R }}
\quad , \cr}
$$
\rightline{(49)}
\noindent
where, for compactness, we have defined
$$
K = f_{\!_E}{ \Gamma w_0 \over{ u (1 + 4u^2 /3) w_1 }} .
\eqno(50)
$$
The matrix ${\cal A}$ then takes the form
$$
{\cal A}=\pmatrix{
0&{a\Gamma/{b\rho w}}&0&0&0\cr\cr
{c_s^2a\rho w/{b\Gamma}}&0&0&0&0\cr\cr
{a\rho/{b\Gamma}}&0&0&0&0\cr\cr
{aw_0({4/3}+f_{\!_E})/{b\Gamma}}&-{2aw_1/{b\rho w}}&0&0&{a/b}\cr\cr
2aw_1(1+K)/b{\Gamma}&-{aw_0}({4/3}+f_{\!_E})/{b\rho w}&0&
{a}({1/3}+f_{\!_E})/b&0\cr} .
\eqno(51)
$$
Next, we introduce $l_i$, the set of left eigenvectors of ${\cal A}$,
and $\lambda_i$, the corresponding eigenvalues satisfying the relations
$$
l_i {\cal A} = \lambda_i l_i .
\eqno(52)
$$
Equation (52) has five distinct eigenvalues (the system is hyperbolic)
$$
\eqalignno{\lambda_0&=0 , &(53)\cr
\lambda_{1,2}&={\pm}\ {a\over b}c_s , &(54)\cr
\lambda_{3,4}&={\pm}\ {a\over b}\sqrt {{1\over 3} + f_{\!_E}} ,  & (55)\cr}
$$

\noindent
to which correspond the five eigenvectors

$$
l_0\equiv k\biggl({0,\;{\pm}\;{1\over {c_s^2w}},\;1,\;0,\;0}\biggr),
\eqno(56)
$$
$$
l_{1,2}\equiv k\biggl({{\pm}\;
{c_s\rho w\over {\Gamma}},\;1,\;0,\;0,\;0}\biggr),
\eqno(57)
$$
$$\displaylines{\qquad{}l_{3,4}\equiv k\biggl({\pm}\;{2(c_s^2-1-K)
(1/3+f_E)^{1/2}\over
{\Gamma(c_s^2 -1/3-f_E)}}w_1+{1\over {\Gamma}}
\biggl({{4\over 3}+f_E}\biggr)w_0,\hfill\cr\cr
\hfill{}{2(f_E-2/3-K)\over {\rho w(c_s^2-1/3-f_E)}} w_1 ,\;0,\;
{\pm}\;\biggl({{1\over 3}+f_E}\biggr)^{1/2},\;1\biggr) .\qquad\cr}
$$
\rightline{(58)}
where $k$ is an arbitrary constant. \par

Equation (46) can then be multiplied on the left by $l_i$ so as to
obtain the symbolic expression
$$
l_i\biggl[{{\partial {\bf U}\over {\partial t}}+
\lambda_i{\partial {\bf U}\over{\partial {\mu}}}}\biggr]+l_i{\bf B}=0 .
\eqno(59)
$$
in which each component involves derivatives only along the
characteristic direction
given by $d \mu / dt = \lambda_i$ . Writing out system (59) explicitly
we then get the following characteristic form of the equations (35) --
(39)
$$
\displaylines{\qquad{} du\;{\pm}\;{\Gamma\over {\rho w c_s}}dp+
a\biggl\{{\Gamma\over {\rho w}}(s_1\;{\pm}\;c_ss_0)+\hfill\cr
\hfill{}+4\pi G R\biggl[{p+\biggl({{1\over 3}+
f_{\!_E}}\biggr)w_0\;{\mp}\;c_sw_1}\biggr]+
{GM\over {R^2}}\;{\pm}\;{2\Gamma uc_s\over R}\biggr\}dt=0, \qquad\cr }
$$
\rightline{(60)}
\medskip
\noindent
which are to be solved along the forward and backward characteristics of
the standard fluid $d\mu = \pm \ (a/b) c_s\,dt$ ,
$$
\eqalign{&dw_1\;{\pm}\;\biggl({{1\over 3}+f_{\!_E}}\biggr)^{\!1/2}dw_0+\cr
&+\biggl[{\biggl({{4\over 3}+f_{\!_E}}\biggr)w_0\;{\pm}
\;{2(c_s^2-1-K)
(1/3+f_{\!_E})^{1/2}\over {c_s^2-1/3-f_{\!_E}}}w_1}\biggr]{1\over
{\Gamma}}du+\cr
&+\biggl[{2(f_{\!_E}-2/3-K)\over
{\rho w(c_s^2-1/3-f_{\!_E})}}\biggr]w_1dp+a\Biggl\{
\biggl({{2u\over R}-{4\pi GRw_1\over {\Gamma}}}\biggr)\times\cr
&\times \biggl[{{2[(1/3+f_{\!_E})(c_s^2-1)-K c_s^2]\over
{c_s^2-1/3-f_{\!_E}}}w_1\;{\pm}\;\biggl({{4\over 3}+f_{\!_E}}\biggr)
\biggl({{1\over 3}+f_{\!_E}}\biggr)^{\!1/2}w_0}\biggr]+\cr
&+\biggl[{4\pi R\biggl({p+w_0\biggl({{1\over 3}+f_{\!_E}}\biggr)}\biggr)+
{M\over {R^2}}}\biggr]\times \cr
&\times \biggl[{\biggl({{4\over 3}+f_{\!_E}}\biggr)w_0\;{\pm}\;
{2(c_s^2-1-K)(1/3+f_{\!_E})^{1/2}\over {c_s^2-1/3-f_{\!_E}}}w_1}\biggr]
{G\over {\Gamma}}+\cr
& \hskip 2.0truecm - {K u (1+4 u^2 /3) \over {\l_{h,q} (1 + R/ \l_{h,q})}}
w_1 + \cr
&-{1\over R}\biggl\{{3f_{\!_E}\biggl[{ {\pm} \biggl({{1\over
3}+f_{\!_E}}\biggr)^{\!1/2}u - \Gamma }\biggr]w_0\;-2\;\biggl[{ {\pm}
\biggl({{1\over 3}+f_{\!_E}}\biggr)^{\!1/2} \Gamma - u}\biggr]w_1}\biggr\}+\cr
&+\biggl[{{2c_s^2\over {\rho w (c_s^2-1/3-f_{\!_E})}}\biggl({f_{\!_E}-{2\over
3}
- K}
\biggr)w_1\;{\mp}\;\biggl({{1\over 3}+f_{\!_E}}\biggr)^{\!1/2}}\biggr]s_0\; +
\cr
& + \;\biggl[{\pm}  {  2 \over {\rho w (c_s^2-1/3-f_{\!_E})}}
\biggl({f_{\!_E}-{2\over 3} - K}\biggr)
\biggl({{1\over 3}+f_{\!_E}}\biggr)^{\!1/2}w_1\; - \;1\biggr] s_1\Biggr\}dt=0 ,
\cr\cr}
$$
\rightline{(61)}
\medskip
\noindent
which are to be solved along the forward and backward characteristics of
the radiation fluid $d\mu = \pm \ (a/b) (1/3 +f_{\!_E})^{1/2}dt$ , and
$$
d\rho-{1\over {c_s^2w}}dp-{as_0\over w}dt=0 ,
\eqno(62)
$$
which is an advective equation and is to be solved along the flowlines
of the standard fluid $d\mu = 0$. Finally, $R$ and $M$ are calculated
from advective equations
$$
dR = a u \, dt ,
\eqno(63)
$$
$$
dM=-4\pi R^2 a u \biggl[{p+\biggl({{1\over 3} +f_{\!_E} }\biggr)w_0
+ {\Gamma\over u}w_1}\biggr]dt ,
\eqno(64)
$$
and the metric coefficient $a$ is calculated from (25) which is a
constraint equation on the constant $t$ hypersurface (i.e. it is to be
integrated along the direction $dt = 0$). \par

The configuration of characteristic curves adjacent to the interface is
shown in Figure 1 for evolution of the system from time level $t$ to a
subsequent time level $t + \Delta t$. The dashed lines represent the
forward and backward characteristics for the radiation fluid {\bf r},
the full narrow lines are the equivalent characteristics for the standard
fluids {\bf f}, the vertical dotted line is the advective characteristic
for strongly interacting matter in the quark phase and the heavy line is
the worldline of the interface.

\medskip
\bigskip
\line{\hskip 1.5 truecm \hfill \psfig{figure=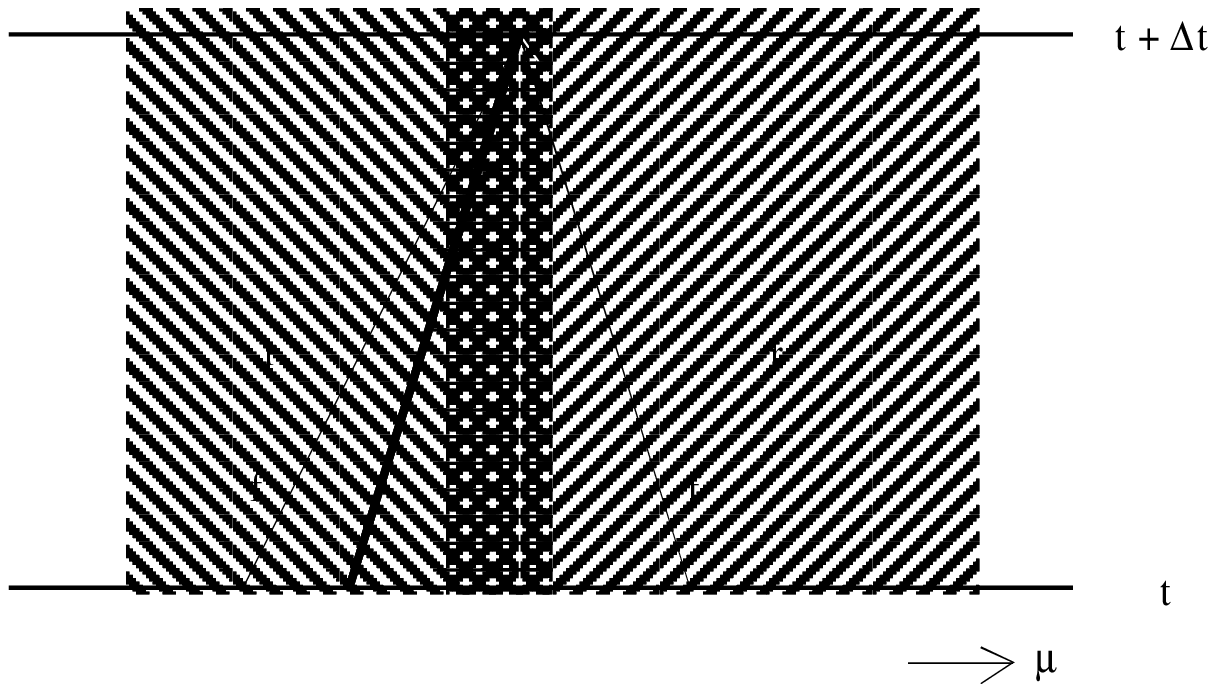,height=8.0truecm
,width=13.0truecm,angle=270} \hfill}

\medskip
\centerline{
\vbox{\hsize=13.0truecm \baselineskip=12pt minus 1pt \noindent\eightrm
Figure 1. The configuration of characteristic curves near the phase
interface drawn in the Lagrangian coordinate frame.}}
\bigskip

The difference between the sound speeds in the standard fluid ($c_s$)
and in the radiation fluid ($(1/3 +f_{\!_E})^{1/2}$) is large when the
former is non-relativistic but, in the present case, the standard fluid
{\it is} relativistic (with $c_s \to 1/\sqrt {3} $) and the difference
between the sound speeds is frequently small. This leads to some serious
complications in numerical solution of the equations which we will be
discussing in paper [11]. \par

The junction conditions linking the values of quantities on either side
of the interface need to take account of the surface tension and surface
energy within it. A complete relativistic treatment of this problem can
be given most conveniently by using the Gauss-Codazzi formalism
[$21,\,22$]. In view of some recent comments [23], it is worth stressing
that this is an economical way to proceed when working within a
relativistic Lagrangian framework even in situations where gravity can
be neglected.
\par

Following the approach described in [9], we can write the following
equations expressing continuity across the interface of $R$, $dR/dt$ and
$ds$
$$
[R]^{\pm}=0 ,
\eqno(65)
$$
$$
[au+b \dot \mu_{_S} \Gamma]^{\pm}=0 ,
\eqno(66)
$$
$$
[a^2-b^2 {\dot \mu_{_S}}^2]^{\pm}=0 ,
\eqno(67)
$$
where $[A]^{\pm} = A^+ -  A^-$, $\{A\}^{\pm} = A^+ + A^-$ and the
superscripts $\pm$ indicate quantities immediately ahead of and behind
the interface. \par

The junction conditions for the energy and momentum of the standard
fluids are
$$
[(e+p)ab]^{\pm}=0 ,
\eqno(68)
$$
$$
[eb^2{\dot \mu_{_S}}^2+pa^2]^{\pm}=-{\sigma f^2\over 2}\biggl\{
{{1\over {ab}}{d\over {dt}}\biggl({b^2\dot \mu_{_S}\over f}\biggr) +
{f_{,\mu}\over {ab}} +
{2\over {fR}}{(b\dot \mu_{_S} u + a\Gamma)}}\biggr\}^{\pm} ,
\eqno(69)
$$
where $\sigma$ is the surface tension (taken here to be independent
of temperature) and $f = (a^2 - b^2 \dot \mu^2_{_S} )^{1/2}$. Since the
thickness of the interface is much smaller than the mean-free-path of
the particles of the radiation fluid, it is reasonable to neglect any
interaction of the latter with the interface, so that the energy and
momentum junction conditions for the radiation fluid simply reduce to
the continuity equations
$$
\biggl[{ab\dot \mu_{_S}\biggl({{4\over 3}+f_{\!_E}}\biggr)
w_0-(a^2+b^2{\dot \mu_{_S}}^2)w_1}\biggr]^{\pm}=0 ,
\eqno(70)
$$
$$
\biggl[{\left\{{a^2\biggl({{1\over 3} + f_{\!_E}}\biggr)+ b^2{\dot \mu_{_S}}^2}
\right\}w_0-2ab\dot \mu_{_S} w_1}\biggr]^{\pm}=0 .
\eqno(71)
$$
The mass function $M$ receives a contribution from the surface energy.
At the time of nucleation of the bubble, conditions are essentially
Newtonian so that
$$
[M]^{\pm}=4\pi R^2 \sigma ,
\eqno(72)
$$
and the subsequent time evolution is given by
$$
{d \over {dt}}[M]^{\pm} = 4\pi R^2 \left[ { b \Gamma {\dot \mu_{_S}}
\left\{ { e + w_0 + {u \over \Gamma}w_1 } \right\} - a u \left\{ { p +
\left( { {1\over 3} + f_{\!_E} } \right) w_0 + {\Gamma \over u} w_1 }
\right\} } \right]^{\pm}
\eqno(73)
$$
Adding conditions (65) -- (73) to equations (14) -- (16), (21) -- (31), and
using the characteristic method (equations (60) -- (64)) to give a
correct treatment of the causal structure, a satisfactory calculation of
the growth of a cosmological hadronic bubble in the presence of long
range energy and momentum transfer can then be made. \par

\bigskip
\bigskip
\noindent
{\bf  7. Numerical Computations of Bubble Growth}
\bigskip

A computer code has been constructed in order to implement the formalism
presented in the previous sections so as to make calculations of the
growth of hadronic bubbles during the cosmological Quark-Hadron
transition. As mentioned previously, the code advances the solution from
one time-level to the next using a standard Lagrangian finite-difference
form of the continuum equations for the bulk of each phase, while the
interface is tracked continuously through the grid using a
characteristic method together with imposition of junction conditions.
This code is a further development of the one described in reference
[10]. A complete account of the numerical techniques used for these new
calculations and of the results obtained will be presented in our
forthcoming paper [11]. \par

\bigskip

\centerline{\psfig{figure=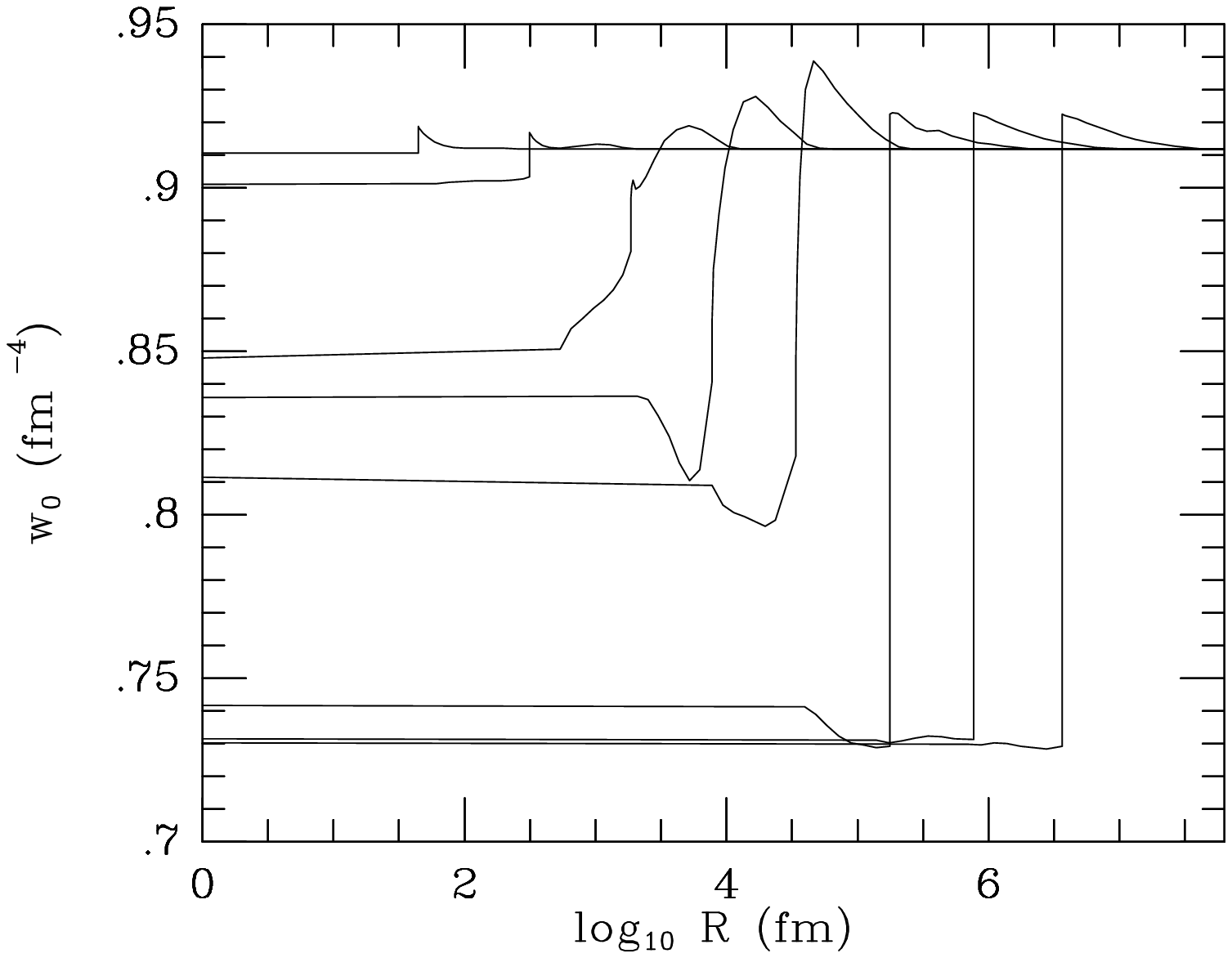,height=11.0truecm,width=13.0truecm}}

\medskip
\centerline{\vbox{
\hsize=13.0truecm \baselineskip=12pt minus 1pt
\eightrm
\textfont1=\eightmit
\scriptfont0=\fiverm
\noindent
Figure 2. Time evolution of the radiation energy density $w_0$. The
sequence of graphs shows the change in the profile of $w_0$ as the bubble
expands. The horizontal dashed line shows the initial profile at the
time of bubble nucleation.}}
\bigskip

As a sample of the results, Figure 2 shows the progressive
coupling together of the radiation fluid and the standard fluid as a
hadronic bubble expands with the mean-free-path of the radiation
particles becoming progressively smaller compared with the radius of the
bubble. Eventually, the coupling becomes essentially total, the
radiative transfer of energy and momentum ceases to operate and the flow
tends towards self-similarity. \par
\bigskip
\bigskip
\noindent
{\bf  8. Conclusion }
\bigskip

Many astrophysical and cosmological situations involve non-stationary
spherical relativistic flows in which a net transfer of energy and
momentum occurs between a radiation fluid and a standard fluid. In order
to describe this kind of flow, we have presented here a new set of
Lagrangian hydrodynamical equations in which the radiative transfer
problem is solved using the PSTF tensor formalism truncated at the
second order. The equations obtained in this way describe the behaviour
of the radiation variables and need to be coupled with the set of
hydrodynamical equations derived from standard conservation laws. \par

One application of particular interest to us concerns bubble growth
during the cosmological Quark--Hadron transition. We have used our
set of equations in a numerical study of this problem, investigating the
effects of energy and momentum transfer by weakly and
electromagnetically interacting particles. In doing this we have treated
the interface between the two phases of the standard fluid as an exact
discontinuity surface and have tracked it continuously through the
finite-difference grid using a characteristic method together with
imposition of junction conditions derived with the aid of the
Gauss-Codazzi equations. The interface moves subsonically and, within
our scheme, use of the characteristic equations is then particularly
important in order to give a correct treatment of the causal structure.
A computer code has been constructed for solving this system of
equations and a full account of the techniques used and results obtained
will be presented in a subsequent paper [11].

The set of equations presented here is general in nature and it can be
applied directly to other situations of interest. Work is now in
progress on applying the equations of Sections 3 and 4 to the study of
non-stationary spherical accretion onto a black hole. A computer code
has been constructed for this and is at present under test.
\vfill\eject
\noindent
{\bf Acknowledgements}
\bigskip
\noindent
We gratefully acknowledge helpful discussions with Ornella Pantano,
Roberto \hbox{Turolla}, Luca Zampieri and Marcello Anile. Financial
support for this research has been provided by the Italian Ministero
dell'Universit\`a e della Ricerca Scientifica e Tecnologica.

\vfill\eject
\noindent
\references
\paper{1} {Thorne K S} {1981} {M.N.R.A.S.} {194}{439}
\paper{2} {Thorne K S, Flammang R A and Zytkow A} {1981} {M.N.R.A.S.}
{194} {475}
\paper{3} {Anile A M and Romano V} {1992} {Ap. J.} {386} {325}
\paper{4} {Nobili L, Turolla R and Zampieri L} {1991} {Ap. J.}
{383}{250}
\paper{5} {Nobili L, Turolla R and Zampieri L} {1993} {Ap. J.}
{404}{686}
\preprint{6} {Zampieri L, Turolla R and Treves A} {1993} {Ap. J. (in
press)}
\paper{7} {Park M-G} {1993} {Astron. Astrophys.} {274}{642}
\paper{8} {Turolla R and Nobili L} {1988} {M.N.R.A.S.} {235}{1273}
\paper{9} {Miller J C and Pantano O} {1989} {Phys. Rev. {\rm D}}
{40}{1789}
\paper{10} {Miller J C and Pantano O} {1990} {Phys. Rev. {\rm D}}
{42}{3334}
\preprint{11} {Miller J C and Rezzolla L} {1993} {(in preparation)}
\paper{12} {Melia F and Zylstra G J} {1991} {Ap.J.}{374}{732}
\paper{13} {Rummukainen K} {1991} {Nucl. Phys. {\rm B} (Proc. Suppl.)}
{24}{47}
\paper{14} {Bonometto S A and Pantano O} {1993} {Physics Reports}
{228}{175}
\paper{15} {Kurki-Suonio H} {1988} {Phys. Rev. {\rm D}} {37} {2104}
\paper{16} {Witten E} {1984} {Phys. Rev. {\rm D}} {30} {272}
\paper{17} {Pantano O} {1989} {Phys. Lett. {\rm B}} {224} {195}
\paper{18} {Chodos A, Jaffe R L, Johnson K, Thorn C B and Weisskopf V F}
{1974} {Phys. Rev. {\rm D}} {9}{3471}
\paper{19} {Gyulassy M, Kajantie K, Kurki-Suonio H and McLerran L} {1984}
{Nucl. Phys. B}{237}{477}
\book{20} {Courant R and Friedrichs K O} {1948} {Supersonic Flow and Shock
Waves} {Springer Verlag, New York}
\paper{21} {Israel W} {1966} {Il Nuovo Cimento} {44}{1}
\paper{22} {Maeda K} {1986} {General Relativity and Gravitation}
{18}{931}
\preprint{23} {Cheng B, Schramm D N and Truran J W} {1993} {Fermilab
preprint 93/093-A}

\vfill\eject

\noindent
{\bf Appendix A}
\bigskip

In this Appendix we give a brief sketch of the calculations leading from
equations (12) -- (13) and (17) -- (20) to the hydrodynamical equations
(15), (16) and (21) -- (29). For clarity, we make a separation into two
sub-sections, one for the radiation fluid alone and the other for the
combined fluids.
\par

\bigskip
\medskip
\centerline {THE RADIATION FLUID}
\bigskip

As mentioned in Section 3, we use the spherically symmetric line element (5)
and expression (4) for the stress-energy tensor of the radiation fluid.
The first three PSTF moments are written as
$$
{\cal M} = w_0 ,
\eqno({\rm A}1)
$$
$$
{\cal M}^{\a }=w_1 e^{\a }_{{\bf {\hat {\it r}}}} ,
\eqno ({\rm A}2)
$$
and
$$
{\cal M}^{\a \b} = w_2 \biggl({e^{\a }_{{\bf \hat {\it r}}}
e^{\b  }_{{\bf \hat {\it r}}}-
{1\over 2} e^{\a }_{{\bf \hat \theta}} e^{\b  }_{{\bf \hat \theta}} -
{1\over 2} e^{\a }_{{\bf \hat \varphi}}
e^{\b  }_{{\bf \hat \varphi}}}\biggr) ,
\eqno ({\rm A}3)
$$
where $w_0$, $w_1$ and $w_2$ can be physically interpreted as explained
in Section 3 and $({\bf e}\sub0hat ,\; {\bf e}\subrhat ,\; {\bf e}_{{\bf
\hat \theta}},\; {\bf e}_{{\bf \hat \varphi}})$  is the orthonormal
tetrad carried by an observer comoving with the standard fluid.
The components of this tetrad are
$$
\eqalignno{
e^{\a }_{{\bf \hat {\it 0}}}&=\biggl({{1\over a},
\;0,\;0,\;0}\biggr) ,&({\rm A}4)\cr
e^{\a }_{{\bf \hat {\it r}}}&=\biggl({0,\;
{1\over b},\;0,\;0}\biggr) ,&({\rm A}5)\cr
e^{\a }_{{\bf \hat \theta}}& = \left( 0,0,{1 \over R},0\right) ,&({\rm A}6)\cr
e^{\a }_{{\bf \hat \varphi}}& = \left( 0,0,0,{1 \over {R sin \theta}}\right) .
&({\rm A}7)\cr}
$$
The comoving observer's four-velocity {\bf u} and four-acceleration {\bf g}
are given by
$$
u^{\a }=\biggl({{1\over a},\;0,\;0,\;0}\biggr) = e^{\a }_{{\bf \hat {\it 0}}},
\eqno({\rm A}8)
$$
and
$$
g^{\a }=\biggl({0,\;{a_{,\mu}\over ab^2},\;0,\;0}\biggr),
\eqno({\rm A}9)
$$
(so that $ e^{\a }_{{\bf \hat {\it r}}}=g^{\a }/g $, where $ g =
(g^{\a }g_{\a })^{1/2} $). The covariant derivative of the radiation
fluid stress energy tensor is
$$
\displaylines{\qquad
T^{\a \b} _{_{R}\ \ ;\b} ={4\over
3}\biggl({\M _{;\b} u^{\a }u^{\b  }+\M u^{\a} _{\; \; ;\b} u^{\b  }+
\M u^{\a} u^{\b  }_{\; \; ;\b} }\biggr)+{1\over 3}\M^{;\a} +\hfill\cr
\hfill+\M^{\a} _{\; \; ;\b} u^{\b  }+\M^{\a} u^{\b  }_{\; \; ;\b} +
\M^{\b  }_{\; \; ;\b} u^{\a} +\M^{\b  }u^{\a} _{\; \; ;\b} +
\M^{\a \b} _{\ \ \; ;\b}  ,
\qquad\cr}
$$
\rightline{(A10)}
where $\Theta=u^{\a  }_{\;\;;\a} $ is the {\it expansion} and $u^{\a }_{\;\;
;\b}  u^{\b  }=g^{\a }$. The contraction of expression (A10) with $u_{\a }$
then gives
$$
\displaylines{\qquad\quad
u_{\a }T^{\a \b} _{_{R}\ \ ;\b} =-\M_{,\b} u^{\b  }-{4\over 3}\M \Theta
-{g^{\b  }\over g}w_{1,\b} +\hfill\cr
\hfill +w_1\biggl({g^{\a }\over g}\biggr)_{\!\! ;\b} u^{\b  }u_{\a }-
w_1\biggl({g^{\b  }\over g}\biggr)_{\!\! ;\b} +\M^{\a \b} _{\ \ \;;\b} u_{\a }
{}.
\qquad\quad\cr}
$$
\rightline{(A11)}
After some further manipulation of expression (A11) and using the equality
$$
\displaylines{\qquad
u_{\a }\M^{\a \b} _{\ \ \; ;\b} =-{(w_2)_{,\b} \over 2}
\biggl({e^{\a }_{{\bf \hat \t}}e^{\b  }_{{\bf \hat \t}}u_{\a }+
 e^{\a }_{{\bf \hat
\v}}e^{\b  }_{{\bf \hat \v}}u_{\a }}\biggr)+\hfill\cr \hfill +
w_2\biggl[{\biggl({g^{\a }\over g}\biggr)_{\!\! ;\b} {g^{\b  }\over g}u_{\a }
-{1\over 2}(e^{\a }_{{\bf \hat \t}}e^{\b  }_{{\bf \hat \t}}+
e^{\a }_{{\bf \hat \v}}
e^{\b  }_{{\bf \hat \v}})_{;\b} u_{\a }}\biggr],\qquad\cr}
$$
\rightline{(A12)}
it is then possible to rewrite equation (12) in the form (15). We
proceed in the same way with the derivation of equation (13). Bearing in
mind that $n_jP^j_{\ \a}=\delta^1_{\ \a}$, it follows from (A10) that
$$
\displaylines{\quad\qquad
n_jP^j_{\ \a}T^{\;\a\b} _{_{R}\ \ ;\b} ={4\over 3}\M g^1+{1\over 3}
\M_{,1}g^{11}+\hfill\cr
\hfill+\M^1_{\ ;\b} u^{\b  }+w_1{g^1\over g}\Theta +\M^{\b  }u^1_{\ ;\b} +
\M^{1\b} _{\ \ \; ;\b}  .\quad\qquad\cr}
$$
\rightline{(A13)}
Writing out each of the terms explicitly and using the expression
$$
\M^{1\b} _{\ \ \; ;\b} ={1\over {b^2}}(w_2)_{,\m }+
{w_2\over {b^2}}\biggl({{a_{,\m}\over a}+
3{R_{,\m}\over R}}\biggr),
\eqno({\rm A}14)
$$
equation (A13) can finally be recast in the form (16). \par
\vfill\eject
\centerline {THE COMBINED FLUIDS}
\bigskip

Expressions for the non-zero components of the total stress-energy
tensor ($T^{\a \b } = T^{\a  \b }_{_R} + T^{\a  \b }_{_F} $ ) can be
calculated using equation (4) and expressions (A4) -- (A7), and this
gives
$$
\eqalignno{
T^{00}& = {1\over {a^2}}(e+w_0) , &({\rm A}15) \cr
T^{01}& = {w_1\over {ab}} , &({\rm A}16) \cr
T^{11}& = {1\over {b^2}} \biggl({p + {w_0\over 3} +
w_2}\biggr) , & ({\rm A}17) \cr
T^{22}& = {1\over {R^2}} \biggl({p + {w_0\over 3} -
{w_2\over 2}}\biggr) , & ({\rm A}18) \cr
T^{33} &= {1\over {R^2 sin ^2 \theta}}  \biggl({p +
{w_0\over 3} - {w_2\over 2}}\biggr) , & ({\rm A}19) \cr}
$$
where it is easy to see the contributions coming from the standard fluid
and the radiation fluid respectively. \par

We will not write down here the expressions for the non-zero Christoffel
symbols and the relevant components of the Ricci tensor, which are
obtained by straightforward but particularly tedious calculations, but
proceed directly to the form of the four independent Einstein equations
$$
\displaylines{
\quad (T^0_{\ 0}) \hfill\cr
8\pi G (e+w_0)R^2 R_{,\m}+{8\pi GbR^2 R_{,t}\ov a} w_1=\bl\{{R
\bl[{\bl({R_{,t}\ov a}\br)^{\!2}-\bl({R_{,\m}\ov
b}\br)^{\!2}+1}\br]}\br\}_{\!,\m}
, \hfill ({\rm A}20) \cr}
$$
$$
\displaylines{
\quad (T^1_{\ 1}) \hfill\cr
8\pi G \bl({p+{w_0\ov 3}+w_2}\br)+{8\pi G a R^2 R_{,\m}\ov b} w_1=-\bl\{{R
\bl[{\bl({R_{,t}\ov a}\br)^{\!2}-
\bl({R_{,\m}\ov b}\br)^{\!2}+1}\br]}\br\}_{\! ,t} ,
\hfill ({\rm A}21) \cr}
$$
$$
\displaylines{
\quad (T^2_{\ 2}\equiv T^3_{\ 3}) \hfill\cr
\qquad 8\pi G \bl({p+e+{4\ov 3}w_0-2 w_2}\br)=
{1\ov {ab}}\bl[{\bl({a_{,\m}\ov b}
\br)_{\! ,\m}-\bl({b_{,t}\ov a}\br)_{\! ,t}}\br]+\hfill\cr
\hfill +{1\ov {R^2}} \bl[{\bl({R_{,t}\ov a}\br)^{\!2}-\bl({R_{,\m}\ov b}
\br)^{\!2}+1}\br] , \qquad\cr}
$$
\rightline{(A22)}
$$
\displaylines{
\quad (T^0_{\ 1}) \hfill\cr \qquad\qquad\qquad\qquad
8\pi G {b\ov a} w_1=-{2\over {a^2 R}} \biggl({ {a_{,\m}\over a}R_{,t}
+ {b_{,t}\over b}
R_{,\m} - R_{,t \m} }\biggr) . \hfill ({\rm A}23) \cr}
$$
(the symbol in brackets identifies the component referred to). Equation
(A23) is a {\it constraint equation} which, in the form
$$
{b_{,t}\ov b}=-{1\ov {R_{,\m}}}\bl({R_{,\m t}-
{a_{,\m}\ov a}R_{,t}-4\pi GabRw_1}\br)
\eqno({\rm A}24)
$$
has been used frequently in the course of the calculations outlined in
this Appendix. In particular, writing out explicitly the time derivative
on the right-hand side of equation (A21) and making use of the (A24),
the expression
$$
\Gamma\Gamma_{,t}={R_{,\m}\ov b}\biggl[{4\pi G R a w_1 - {R_{,t}\ov b}
\biggl({p_{,\m}+bs_1\ov {e+p}}\biggr)}\biggr],
\eqno({\rm A}25)
$$
can be obtained and this can then be further transformed, by means of
equation (27), so as to arrive at the form (21). \par

Next we turn to writing out explicitly the hydrodynamic conservation
equations (18) -- (20) which then take the form
$$
\eqalignno{
-u_{\a }T^{\a \b} _{\ \ \ ;\b} &=0=s_0+(e+p)_{,\b} u^{\b  }+(e+p)\Theta -
p_{,\b} g^{\a \b} u_{\a }\ \ , &({\rm A}26) \cr
n_jP^j_{\ \a}T^{\a \b} _{\ \ \ ;\b} &=0
={1\over b}s_1+(e+p)n_ju^j_{\;;\b} u^{\b  }+
(g^{\a \b} +u^{\a }u^{\b  })p_{,\b} \ \ , &({\rm A}27) \cr
(\rho u^{\a })_{; \a }&=0=\rho_{, \a } u^{\a } + \rho u^{\a }_{\ \ ,\a}+
\rho u^{\a }{\left( {\sqrt{{\rm det}(-g^{\a \b} )}}\right) _{\!,\a}\over
{\sqrt{{\rm det}(-g^{\a \b} )}}} . &({\rm A}28) \cr }
$$
Using the expressions $\Gamma=R_{,\m}/b$ and $u=R_{,t}/a$ together with (A24),
equations (A26) -- (A28) can be converted to the final form given in
equations (23) -- (25). Finally, we note that if we rewrite equations
(A20) and (A21) as
$$
\bl\{{{R\over{2G}}\bl[{\bl({R_{,t}\ov a}\br)^{\!2}-
\bl({R_{,\m}\ov b}\br)^{\!2}+1}\br]}\br\}_{\! ,\m}=
4\pi R^2 R_{,\m} (e+w_0 + {u \over \Gamma} w_1) = M_{,\mu} ,
\eqno({\rm A}29)
$$
and
$$
\bl\{{{R\over{2G}}\bl[{\bl({R_{,t}\ov a}\br)^{\!2}-
\bl({R_{,\m}\ov b}\br)^{\!2}+1}\br]}\br\}_{\! ,t}=
- 4\pi R^2 R_{,t} \bl({p+{1\ov 3} w_0+{\Gamma \over u}w_1+w_2}\br) =
M_{,t} ,
\eqno({\rm A}30)
$$
this gives expressions (26) and (29) for the generalized mass function
$M$. These reduce to the familiar expressions for the standard mass
function when the radiation terms are omitted.
\par

\vfill\eject

\noindent
{\bf Appendix B}
\bigskip

As mentioned in Section 3, a particularly important point in the PSTF
approach is the truncation of the infinite hierarchy of moment equations
by means of a suitably defined closure relation derived on the basis of
physical considerations. We have used the closure equation (14) and, for
our particular problem, an appropriate expression for $f_{\!_E}$ is given by
equation (47). This appendix gives our justification for making this
choice.
\par

A key point in choosing an expression for $f_{\!_E}$ is that it should
have the correct asymptotic behaviour in the optically-thin and
optically-thick limits. We are considering the case of a single
spherical hadronic bubble which is initially nucleated at rest with a
radius small compared with the mean-free-path of the radiation. Under
these circumstances, the radiation field will be everywhere rather
accurately uniform and isotropic (unless there is some other perturbing
influence) and since the bubble radius is very small compared with the
horizon scale, it is also a good approximation to neglect cosmological
expansion. Since $w_0$, $w_1$ and $w_2$ are all measured with respect to
the local rest-frames of the standard fluid, the values which they take
during the early part of the bubble expansion are those produced by
motion of the fluid rest frames with respect to the essentially uniform
radiation field.
\par

When the radiation is isotropic in its mean rest frame, its
stress-energy tensor takes the perfect fluid form
$$
T^{\a  \b }_{_R} = (e_{_R} + p_{_R}) v^{\a } v^{\b } + p_{_R} g^{\a  \b } ,
\eqno({\rm B}1)
$$
where $e_{_R}$ and $p_{_R}$ are the radiation energy density and pressure
($p_{_R} = e_{_R}/3$), measured in the mean rest frame of the radiation,
and $v^{\a } = d x^{\a }/d\tau$ is the four-velocity of this frame
relative to some specific observer. For purely radial motion in our
metric
$$
d\tau^2=-ds^2=\biggl({1-{b^2{\dot \m}^2\over {a^2}}}\biggr)a^2dt^2=
{a^2 \over {\gamma^2}} dt^2 ,
\eqno({\rm B}2)
$$
and the non-zero components of four-velocity are
$$
v^{\a}=\biggl({{dt\over {d\tau}},{d\m\over {d\tau}}}\biggr)=
{\gamma\over a}(1,{\dot \m}) ,
\eqno({\rm B}3)
$$
where
$$
\gamma={1\over {\sqrt{1-v^2}}}  \hskip 2.0truecm {\rm and} \hskip 2.0truecm
v={b{\dot \m}\over {a}}.
\eqno({\rm B}4)
$$
To find the value of $\dot \m$ for the radiation frame with respect to
the standard fluid, we note that since each element of the radiation fluid
is remaining at a constant value of $R$
$$
dR = R_{,t} dt + R_{,\mu} d\mu = au dt + b \Gamma d\mu =0 ,
\eqno({\rm B}5)
$$
and so
$$
{\dot \m}=-{au \over {b \Gamma}} ,
\eqno({\rm B}6)
$$
which leads to the following expressions
$$
v=-{u \over {\Gamma}} \hskip 5.0truecm
 \gamma^2={\Gamma^2 \over {\Gamma^2 - u^2}} .
\eqno({\rm B}7)
$$
If we now write the stress-energy tensor (B1) in the frame comoving with
the standard fluid, we can compare the new expressions for the
components with the ones appearing in equation (4) and so obtain the
following system of equations
$$
\eqalignno{
T^{00}_{_R}&={w_0 \over {a^2}}
={4\over 3}e_{_R}\left({{\gamma^2} \over {a^2}}\right)
+{1\over 3}e_{_R}\left(-{1 \over {a^2}}\right),
&({\rm B}8)\cr
T^{01}_{_R}&={{w_1} \over {ab}}
={4\over 3}e_{_R} \left({{\gamma^2} \over {a^2}}{\dot \m}\right),
&({\rm B}9)\cr
T^{11}_{_R}&={1 \over {b^2}}\biggl({{w_0\over 3}+w_2}\biggr)=
{4\over 3}e_{_R} \left({{\gamma^2} \over {a^2}}{\dot \m}^2 \right)
+{1 \over 3} e_{_R} \left({1 \over {b^2}}\right).
&({\rm B}10)\cr
}
$$
The solution of this system then leads to the expressions
$$
\eqalignno{
w_0 &= {\gamma^2 \over 3} (3 + v^2) e_{_R}, & ({\rm B}11) \cr
w_1 &= {4 \over 3} \gamma^2 v e_{_R}, & ({\rm B}12) \cr
w_2 &= {\gamma^2 \over 3} (1 + 3v^2) e_{_R} - {w_0 \over 3} . & ({\rm B}13)
\cr}
$$
If we now define $(w_0)_{_N} = e_{_R}$ to be the radiation energy
density at the bubble nucleation time (when there is no fluid motion),
equations (B11) -- (B13) can be suitably transformed so as to give
expressions for the energy density, flux and shear scalar of the
radiation as seen from the standard fluid
$$
w_0=\biggl({1+{4\over 3}{u^2 \over {\Gamma^2-u^2}}}\biggr)(w_0)_{_N}
\simeq \biggl({1+{4\over 3}u^2}\biggr)(w_0)_{_N} ,
\eqno({\rm B}14)
$$
$$
w_1=-{4\over 3} {\Gamma u \over {\Gamma^2-u^2}} (w_0)_{_N}
\simeq -{4\over 3} {\Gamma u } (w_0)_{_N} ,
\eqno({\rm B}15)
$$
$$
w_2={8\over 9}{u^2 \over {\Gamma^2-u^2}} (w_0)_{_N}
\simeq {8\over 9}{u^2 } (w_0)_{_N} .
\eqno({\rm B}16)
$$
The approximate forms of the expressions (B14) -- (B16) result from
noting that since the dimensions of the bubble are small compared with
the horizon scale, $\Gamma^2 - u^2 \simeq 1$. \par

Equations (B14) and (B16) together with the definition (14), give the
following analytic expression for the variable Eddington factor during
the first stages of the bubble expansion
$$
f_{\!_E} = {8 u^2 / 9 \over {(1 + 4 u^2/3)}} .
\eqno({\rm B}17)
$$
This is the ``optically thin'' limit. At the other extreme, the
``optically thick'' limit arises when the radius of the bubble is large
compared with the radiation mean-free-path and complete coupling has been
attained between the radiation and the standard fluid over length-scales
comparable with the radius of the bubble. When this happens,
interactions make the radiation isotropic in the local fluid rest frame
so that $w_2 \to 0$ and $f_{\!_E} \to 0$. A suitable smooth join is
required in between the two asymptotic limits and to do this we have
multiplied the expression in (B17) by $\lambda_{h,q} / (\lambda_{h,q} +
R)$ which then gives equation (47). Experiment has shown that reasonable
variation in the form of the join makes an insignificant change in the
results obtained. \par

\vfill\eject\end